\documentclass[amsfonts, amssymb, amsmath, reprint, showkeys, nofootinbib, twoside]{revtex4-1}
\usepackage[english]{babel}
\usepackage[utf8]{inputenc}
\usepackage[colorinlistoftodos, color=green!40, prependcaption]{todonotes}
\usepackage{amsthm}
\usepackage{mathtools}
\usepackage{physics}
\usepackage{xcolor}
\usepackage{graphicx}
\usepackage{adjustbox}
\usepackage{placeins}
\usepackage[T1]{fontenc}
\usepackage{lipsum}
\usepackage[pdftex, pdftitle={Article}, pdfauthor={Author},colorlinks=false]{hyperref} 
\begin{document}

\title{Long-range Bloch Surface Waves in Photonic Crystal Ridges}

\author{Tommaso Perani}
\email{tommaso.perani01@ateneopv.it}
\author{Marco Liscidini}
\affiliation{Dipartimento di Fisica, Università di Pavia, Via Agostino Bassi 6, 27100 Pavia (PV), Italy}

\date{\today} 

\begin{abstract}
We theoretically study light propagation in guided Bloch surface waves (BSWs) supported by photonic crystal ridges. We demonstrate that low propagation losses can be achieved just by a proper design of the multilayer to obtain photonic band gaps for both light polarizations. We present a design strategy based on a Fourier analysis that allows one to obtain intrinsic losses as low as 5 dB/km for a structure operating in the visible spectral range. These results clarify the limiting factors to light propagation in guided BSWs and represent a fundamental step towards the development of BSW-based integrated optical platforms.
\end{abstract}


\maketitle


Bloch surface waves (BSWs) are electromagnetic waves propagating along the interface between a truncated periodic dielectric multilayer, i.e., a semi-infinite one-dimensional (1D) photonic crystal (PhC), and a homogeneous dielectric medium. Their confinement relies on total internal reflection (TIR) from the homogeneous medium and on the presence of a photonic band gap (PBG) from the multilayer \cite{Yeh:77}. BSWs are an interesting approach to light confinement near the structure surface, with surface field enhancements up to 40\% larger than those achievable in dielectric slabs \cite{Aurelio:17}. In addition, the hybrid confinement mechanism gives more freedom in tailoring the dispersion relations, with the effective index no longer limited by that of the structure substrate \cite{Liscidini:09}. Finally, structures supporting BSWs can be realized in various material platforms from oxides \cite{Robertson:99} to polymers \cite{Frezza:11}, which allows one to operate in a wide spectral range, from mid-infrared to visible wavelengths.

In recent years, several works have focused on the control and manipulation of guided BSWs using dielectric loads on the multilayer surface to establish a complete integrated optical platform based on PhC ridges \cite{Dubey:16,Rodriguez:19,Stella:19,Perani:19}. Unfortunately, so far experimental works have shown propagation losses of the order of dB/mm \cite{Descrovi:10,Dubey:17,Wang:2017}. This is not only frustrating, as it hinders the development of an integrated optical platform based on BSWs, but also unclear from the physical standpoint. Indeed, finite-difference time-domain (FDTD) simulations of idealized structures confirm that such high losses are not due to fabrication imperfections \cite{Rodriguez:19}, but they are intrinsically related to the kind of light confinement that relies on a combination of TIR and PBG.
 
Little research has been conducted into understanding the leading loss mechanisms. Earlier theoretical investigations on BSW dispersion and propagation rely on the use of approximated strategies, mainly based on effective index methods \cite{Menotti:15}. These approaches offer an advantage in terms of computational time and resources to calculating the dispersion relation and field distribution, but at the same time are not able to capture all the physics of light propagation, especially for what concerns propagation losses. In this Letter, by neglecting material absorption and scattering, we explain the physical mechanism behind the intrinsic losses of guided BSWs in PhC ridges and present a strategy to achieve a significant improvement of the mode propagation lengths.

\begin{figure}[!b]
\centering 
{\includegraphics[width=.9\linewidth]{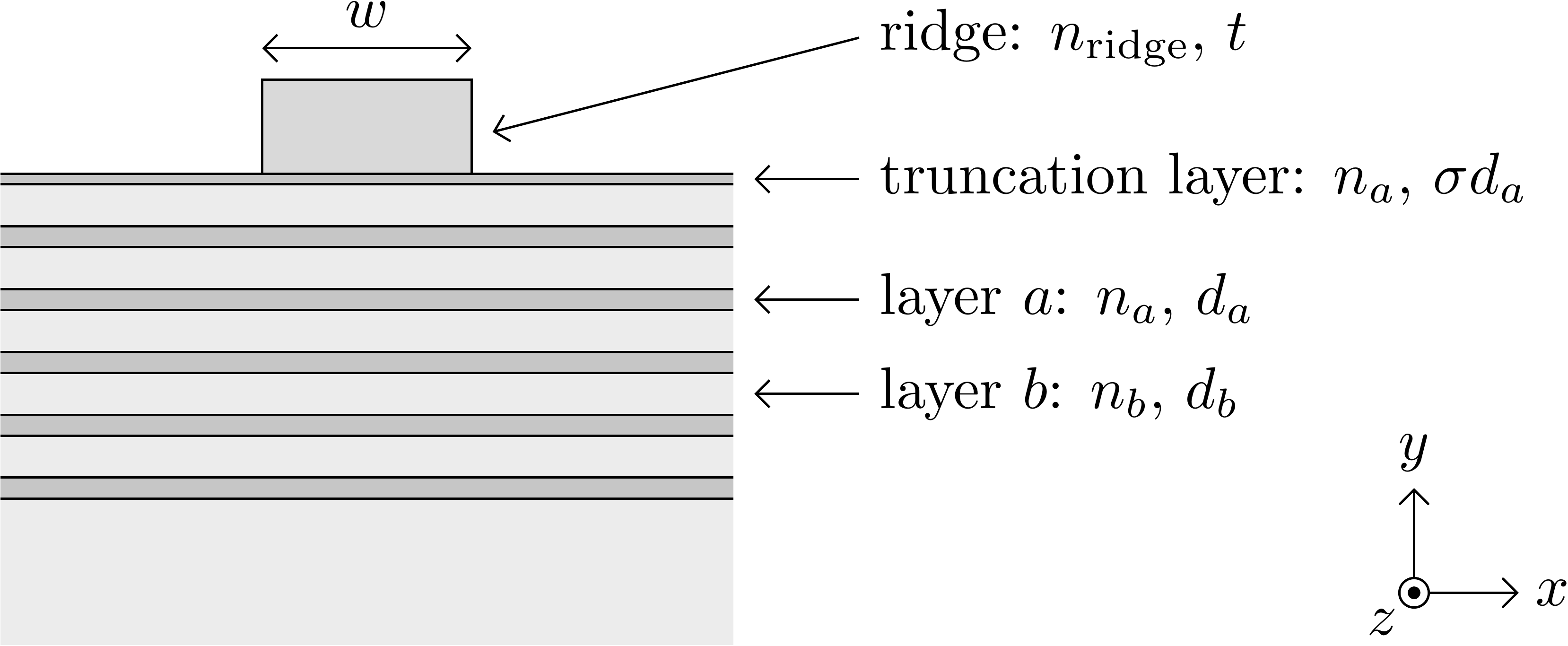}}
\caption{Sketch of the cross section of the PhC ridge.}
\label{fig:ridge}
\end{figure}

The structure under investigation is illustrated in Fig.~\ref{fig:ridge}. It consists of a dielectric ridge placed on the top of a truncated periodic multilayer. The multilayer has a finite number $N$ of periods composed of two alternating layers of thicknesses $d_a$ and $d_b$ stacked along the $y$ direction, with refractive indices $n_a$ and $n_b<n_a$, respectively. The filling fraction $f=d_a/(d_a+d_b)$ determines the spectral position and extension of the (polarization-dependent) PBG with respect to the given refractive indices of the layers \cite{Joannopoulos:08}. The topmost $n_a$-index layer is truncated with thickness $d_\sigma=\sigma d_a$, with $0<\sigma<1$. On the top of the structure, a homogeneous dielectric ridge of width $w$, thickness $t$, and refractive index $n_\text{ridge}$, sustains a guided BSW. The surrounding cladding is air ($n_\text{clad}=1$). The structure is designed to operate at a given wavelength $\lambda_0$ in vacuum. We stress that we are considering a very general case, in which the guided BSW can be either a perturbation of the unguided BSW supported by the bare structure or not. The latter is a special case in which the effective index approaches typically adopted in describing guided BSWs cannot always be applied \cite{Liscidini:12}.

Due to the hybrid combination of TIR and reflection within the PBG, the properties of the guided modes depend on the multilayer and the ridge parameters, which provide great freedom in terms of structure design, and yet sum up to a vast parameter space. However, it is worth reminding that the PBG is a bulk property of the sole multilayer, and its features depend only on its unit cell composition. The multilayer is the main focus of this work, for it is associated with the most complex confinement mechanisms and, as we shall see, its optimization is crucial in reducing the propagation losses.

Even in the presence of isotropic materials, the multilayer response at finite $k_z$ depends on the light polarization. The PBG is intuitively understood to arise from the  interference of the light reflected at each multilayer interface, whose amplitude depends on its polarization according to Fresnel’s coefficients. As a result of the mirror symmetry with respect to the $yz$ plane and any other plane parallel to it, optical modes inside a 1D PhC can be labeled as either transverse-electric (TE) or transverse-magnetic (TM), with either $\bf{E}$ or $\bf{H}$ perpendicular to the $yz$ plane. Consequently, one can classify eventual unguided BSWs supported by a bare truncated multilayer as TE-polarized, with only $E_x,H_y,H_z$ nonvanishing field components, or TM-polarized, with only $H_x,E_y,E_z$ nonzero.

\begin{figure}[!t]
\centering 
\includegraphics[width=\linewidth]{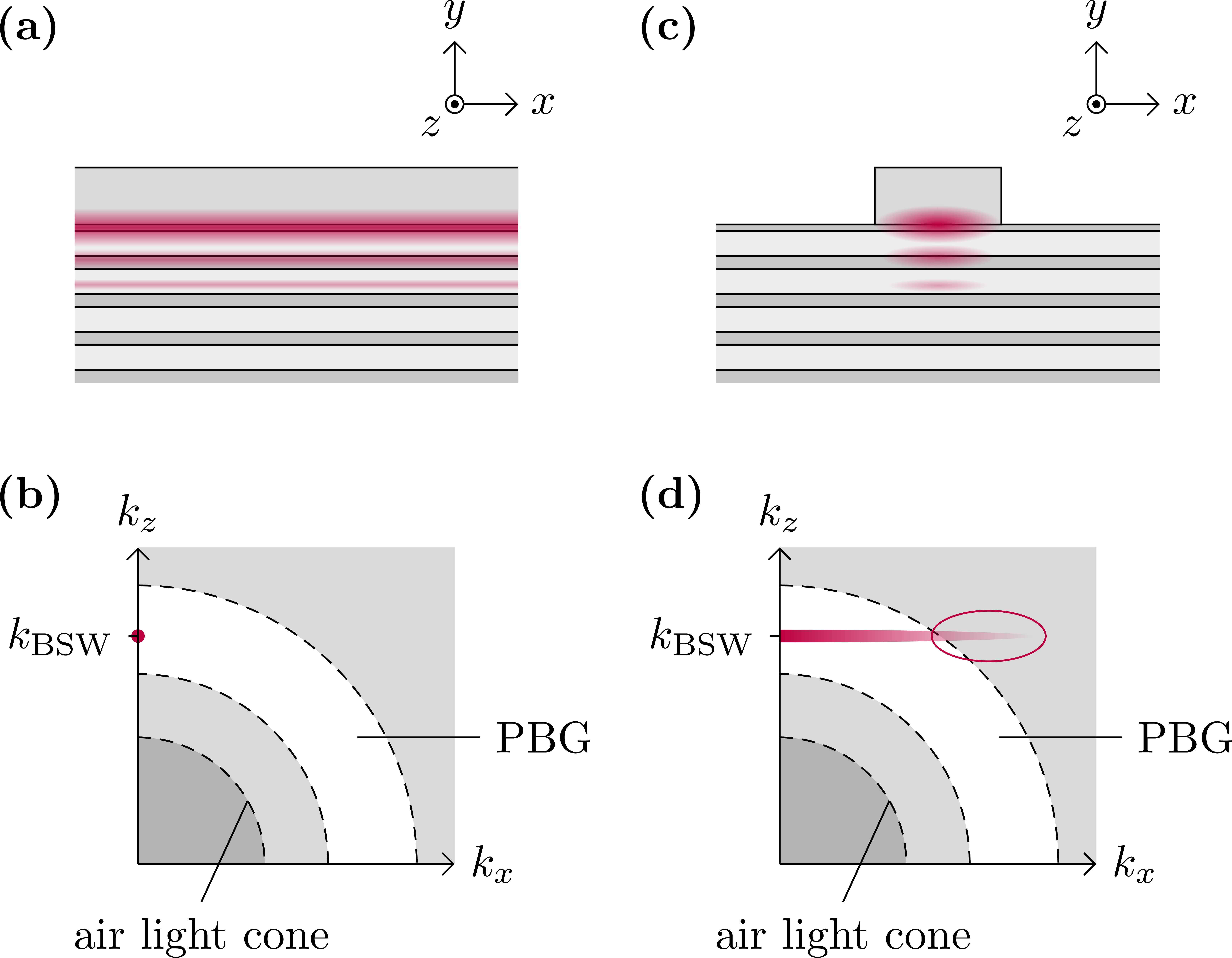}
\caption{Pictorial representation of the field distribution and its Fourier spectrum in the $(k_x,k_z)$ plane of the $E_x$ component for a TE-polarized BSW in a 1D structure (a, b) and for a TE-like guided BSW in a PhC ridge (c, d).}
\label{fig:argument}
\end{figure}

On the contrary, the guided mode propagating in the $z$ direction along the ridge shown in Fig.~\ref{fig:ridge} is  characterized by the propagation constant $k_z=k_\text{BSW}=2\pi n_\text{BSW}/\lambda_0$, with $n_\text{BSW}$ the mode effective refractive index, and the electric and magnetic field profiles ${\bf E}(x,y)$ and ${\bf H}(x,y)$, which now are functions of both transverse spatial coordinates. Due to the finite width of the ridge, the structure is symmetric upon reflection with respect to the sole $yz$ plane, i.e., at the center of the ridge ($x=0$). Thus, in general, for $x\neq0$ all the six field components of any guided mode are nonvanishing. This can also be easily understood by considering that light confinement in the $x$ direction implies $\partial/\partial x \neq 0$. Thus, a purely TE mode ${\bf E}=(E_x,0,0)$ cannot be a solution to the first Maxwell’s equation $\nabla\cdot{\bf D} = 0$, with ${\bf D}({\bf r})=\varepsilon({\bf r}){\bf E}(\bf{r})$. In this situation, guided modes can sometimes be labeled as  TE(TM)-like, in that $E_x$ ($H_x$) is the dominant field component. This is a major feature of any guided mode in two-dimensional (2D) waveguides, and it is of fundamental importance in addressing the origin of propagation losses of guided BSWs. Indeed, while the condition of TIR is independent of light polarization (unless one deals with anisotropic media), in PhC ridges light confinement from the multilayer side is due to a PBG. Thus, the simultaneous presence of TE and TM PBGs around $k_\text{BSW}$ is always required to ensure that all six nonvanishing field components are guided.

A second important aspect is associated with the field distribution in the structure, which is connected to the Fourier components of the mode. In Fig.~\ref{fig:argument}(a) we sketch the case of a purely 1D system, in which light is confined only along the $y$ direction. At a given $\lambda_0$, the mode is characterized by a single point in the Fourier plane $(k_x,k_z)$ (see Fig.~\ref{fig:argument}(b)). On the contrary, in the case of a guided BSW, light confinement along the $x$ direction (see Fig.~\ref{fig:argument}(c)) determines a spreading of the mode Fourier components along $k_x$ (see Fig.~\ref{fig:argument}(d)). Naturally, the tighter the confinement is, the larger the spreading of the field Fourier components is along $k_x$, with the risk of having a significant fraction of them outside the PBG, as illustrated in Fig.~\ref{fig:argument}(d). In this case, light can couple to the radiation modes supported by the multilayer.

\begin{figure}[!t]
\centering 
{\includegraphics[width=.95\linewidth]{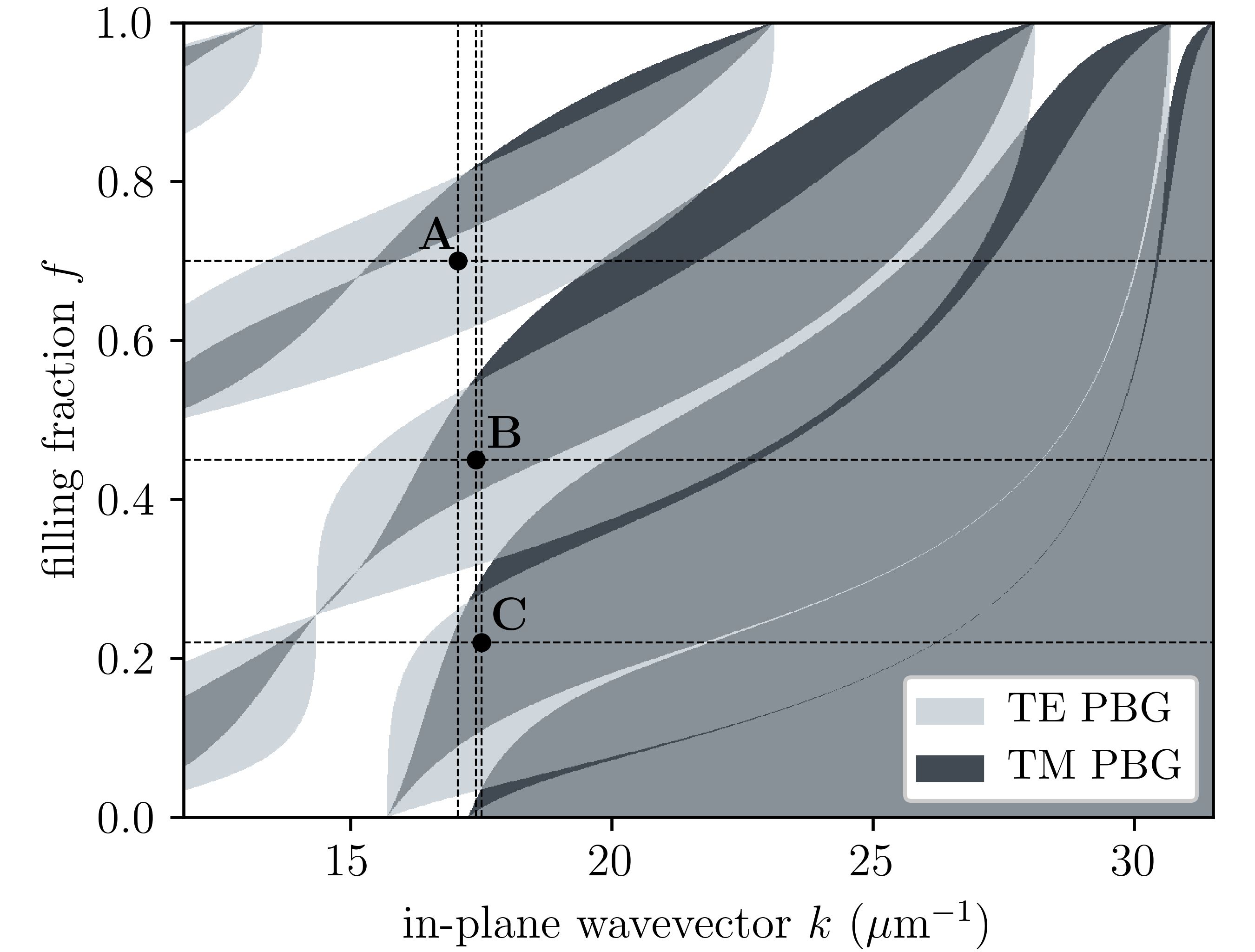}}
\caption{Gap map (outside the air light cone) at $\lambda_0=532\,\text{nm}$ of the present TiO$_2$/SiO$_2$ multilayer of period $\Lambda=440\,\text{nm}$, as a function of the filling fraction $f$ and the in-plane wavevector $k$. The TE (TM) PBG is indicated by the light (dark) area. The three investigated cases are pointed out (dashed lines).}
\label{fig:gapmap}
\end{figure}

For a given wavelength $\lambda_0$ and a set of materials, one can design the structure on the basis of these two arguments to minimize the propagation losses. The idea is that, for a sufficiently large number of periods, longer propagation lengths require that all the field Fourier components are inside a PBG, to avoid leakage into the multilayer. In addition, since light is confined also by TIR, the Fourier components must be outside the cladding light cone, to prevent coupling with the continuum of radiation modes in the cladding.

Following this approach, we consider the case of a PhC ridge realized on a titania (TiO$_2$, $n_a=2.67$)/silica (SiO$_2$, $n_b=1.46$) multilayer of $N=10$ periods $\Lambda=d_a+d_b=440\,\text{nm}$. The topmost TiO$_2$ layer is truncated with thickness $d_\sigma=10\,\text{nm}$ and loaded with a polymethylmethacrylate (PMMA, $n_\text{ridge}=1.49$) ridge of thickness $t=0.4\,\mu\text{m}$ and width $w=1\,\mu\text{m}$ to work in the visible spectrum at $\lambda_0=532\,\text{nm}$. To restrict the number of possible configurations in the parameter space, we first let the filling fraction $f$ be the only free parameter, with the TiO$_2$ and SiO$_2$ layers having thicknesses $d_a=f \Lambda$ and $d_b=(1-f)\Lambda$, respectively. For such a structure, in Fig.~\ref{fig:gapmap} we report  the gap map for both polarizations as a function of the filling fraction $f$ and the modulus $k=\sqrt{k_x^2+k_z^2}$ of the in-plane wavevector $(k_x,k_z)$.

We can identify three different points A, B, and C that represent three qualitatively different situations given the ridge parameters indicated above. The point A corresponds to the case in which the PhC ridge is designed with a filling fraction $f_\text{A}=0.7$ and supports a TE-like BSW with a leading Fourier component at $k=k_\text{BSW,\,A}=17.1\,\mu\text{m}^{-1}$. In this case, one has a strong PBG for the TE components, but no gap for the TM components. The point B, at $f_\text{B}=0.45$ and $k_\text{BSW,\,B}=17.4\,\mu\text{m}^{-1}$, represents the intermediate situation of a strong TE PBG overlapping with a moderately wide TM PBG. Finally, the point C, at $f_\text{C}=0.22$ and $k_\text{BSW,\,C}=17.5\,\mu\text{m}^{-1}$, corresponds to the most favorable case in which both TE and TM PBGs are wide and well-overlapping.

\begin{figure}[!t]
\centering 
{\includegraphics[width=\linewidth]{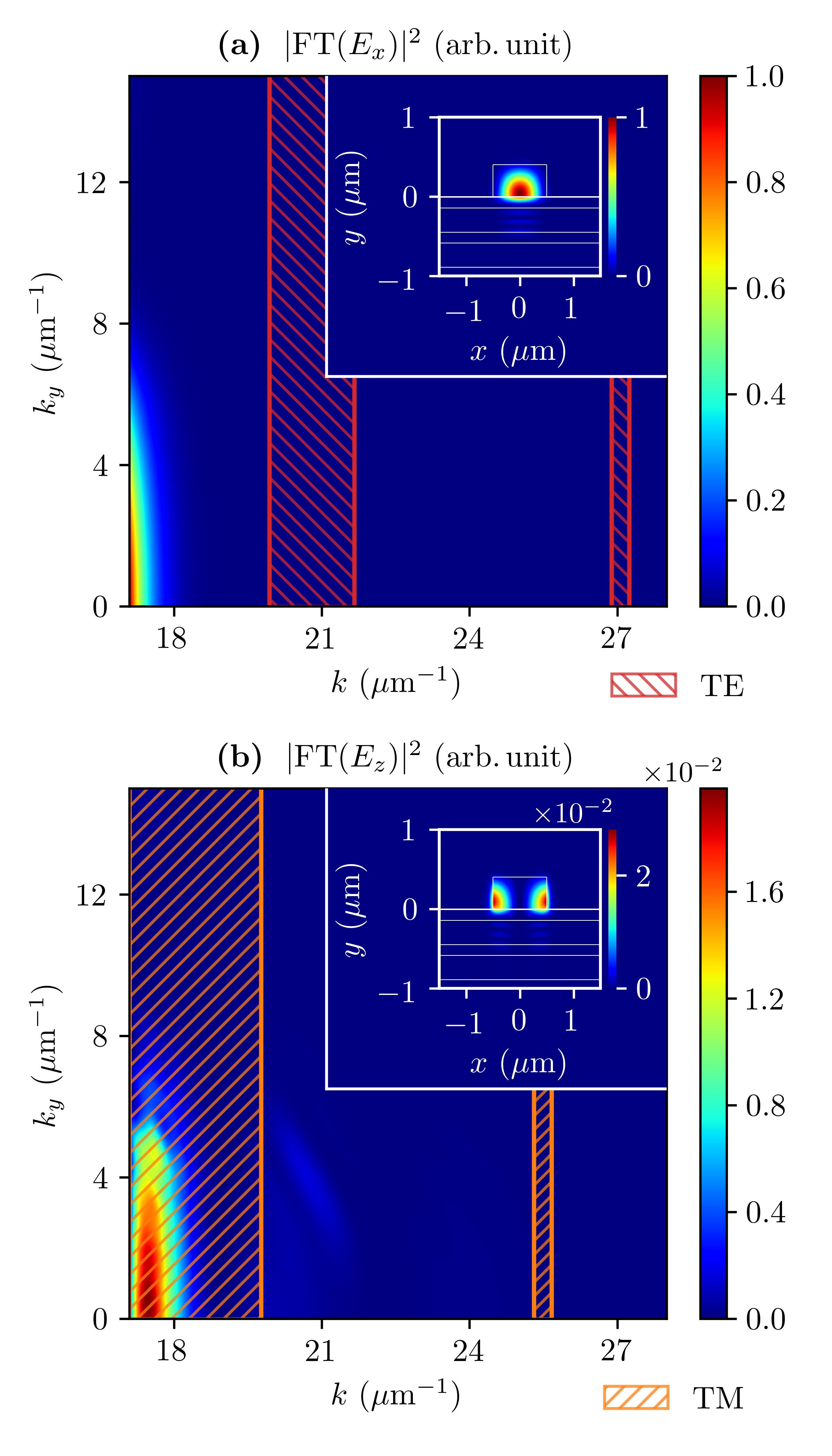}}
\caption{Fourier spectra and intensity profiles (inset), normalized to the maximum value, of the FDE-simulated field components $E_x$ (a) and $E_z$ (b) for the structure corresponding to case A. The dashed areas correspond to values of $k$ for which the TE (a; red) and TM (b; orange) PBGs are closed.}
\label{fig:caseA}
\end{figure}

\begin{figure}[!t]
\centering 
{\includegraphics[width=\linewidth]{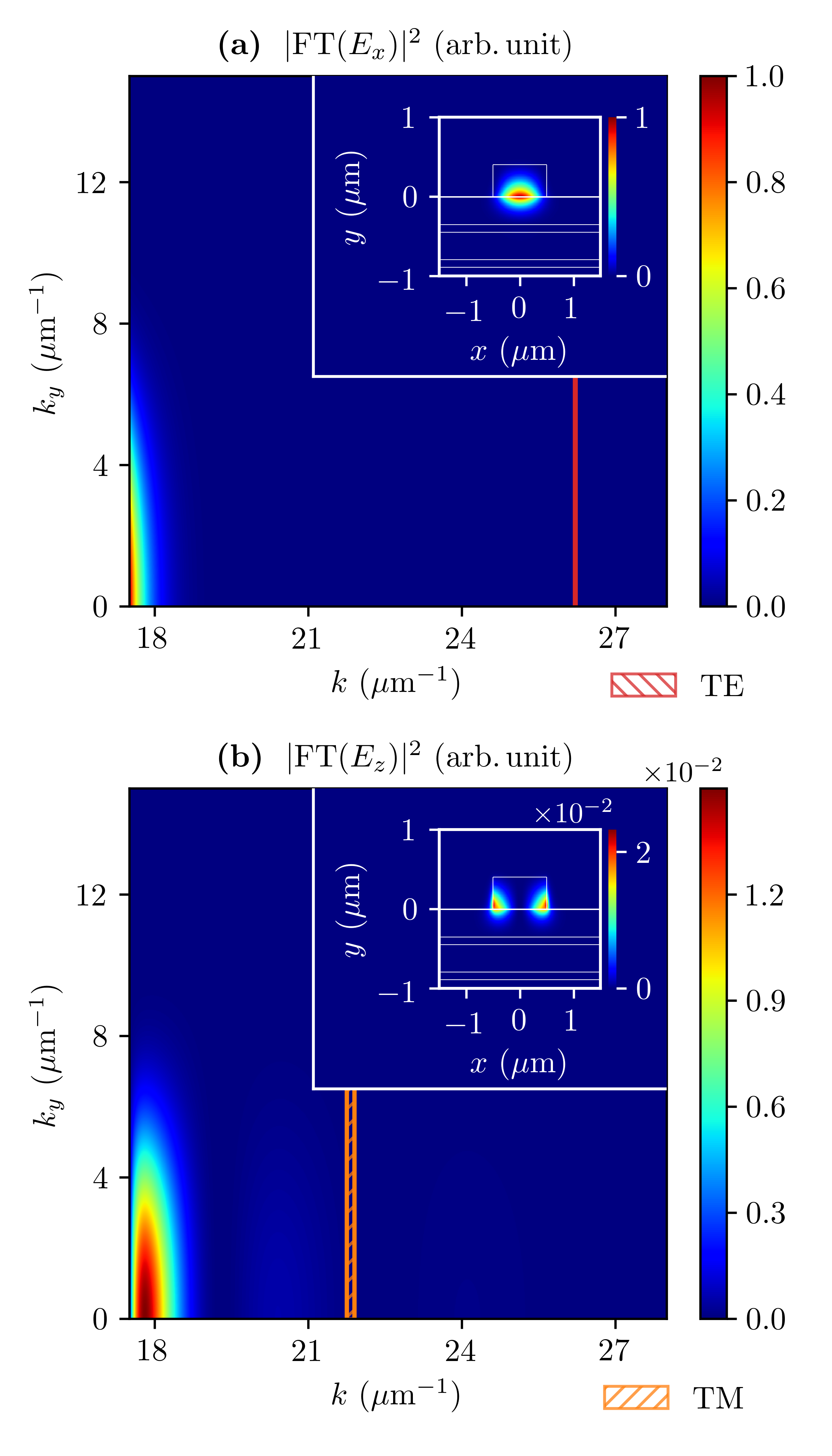}}
\caption{Fourier spectra and intensity profiles (inset), normalized to the maximum value, of the FDE-simulated field components $E_x$ (a) and $E_z$ (b) for the structure corresponding to case C. The dashed areas correspond to values of $k$ for which the TE (a; red) and TM (b; orange) PBGs are closed.}
\label{fig:caseC}
\end{figure}

Following the argument presented in Fig.~\ref{fig:argument}, we calculate the Fourier transform (FT) of each field component as a function of $k$ for the guided modes supported by the PhC ridges A, B, and C. This can be obtained from the electric field profile ${\bf E}(x,y)$ of the modes, which can be computed numerically by means of a 2D finite difference eigenmode (FDE) solver with perfectly matched layer (PML) boundary conditions \cite{Lum}.
It should be noticed that, in general, ${\bf E}(x,y)$ is not separable as a product of two functions in the $x$ and $y$ coordinates separately, thus also the FT is not separable in $k_x$ and $k_y$. This means that, for a proper analysis, one has to consider the FT distribution as a function of both $k$ and $k_y$, as plotted in Figs.~\ref{fig:caseA} and \ref{fig:caseC}.

We start by considering the case $f_\text{A}=0.7$ (working point A of Fig.~\ref{fig:gapmap}) for which $k_\text{BSW,\,A}=17.1\,\mu\text{m}^{-1}$. The Fourier spectra as a function of $k$ and $k_y$ for the two field components $E_x$ and $E_z$ are shown in Figs.~\ref{fig:caseA}(a) and  (b), respectively, along with the corresponding real-space intensity profiles (inset). The results for $E_y$ are not shown because its field intensity is sufficiently small to be neglected. In each graph, the dashed regions correspond to values of $k$ for which the PBG for the relevant polarization is closed, and light is not guided. We observe that, while almost the entire FT for $E_x$ is within a PBG, this is not so for $E_z$, for which most of the light cannot be confined by the multilayer. It is confirmed by the computed propagation losses that are $\alpha_\text{A}=50.0\,\text{dB/cm}$ \cite{loss}.

A similar analysis can be done for the structure $f_\text{B}=0.45$ (working point B of Fig.~\ref{fig:gapmap}), for which there exist TE and TM PBGs around the working point. In this case, the Fourier analysis (not shown) unveils that a significant fraction of the Fourier components is within the TE and TM PBGs around $k_\text{BSW,\,B}=17.4\,\mu\text{m}^{-1}$. The previous example suggests that in this case lower propagation losses should be expected. Indeed, the computed value is $\alpha_\text{B}=1.03\,\text{dB/cm}$. This improvement is of more than one order of magnitude and indicates that a further optimization of the PBG position to confine the TM fraction of the mode could lead to better results.

With this in mind, we now turn to the case $f_\text{C}=0.22$ (working point C of Fig.~\ref{fig:gapmap}). This point represents a favorable situation of wide and well-overlapping TE and TM PBGs. In this case, the computed propagation losses are $\alpha_\text{C}=5.67\times 10^{-5}\,\text{dB/cm}$, six orders of magnitude lower than those of case A. The origin of such a noteworthy improvement is clear by looking at Fig.~\ref{fig:caseC}, where we show the Fourier analysis along with the electric field intensity profiles for the two dominant field components. In this case, almost all the field FTs are within the PBG regions for both polarizations. All the results are summarized in Tab.~\ref{tab:cases}.

So far, we focused on the multilayer parameters that control the position and extension of the PBGs. However, one can also modify the FT distribution of the guided BSW by varying the ridge width $w$. This gives an additional tuning parameter that can be particularly useful in practice, when the multilayer is already fabricated and one wants to improve the propagation lengths even further. In the limit of wide ridges, propagation losses are expected to decrease, for the mode is more localized in the Fourier space and its polarization tends to become purely TE or TM. In the more interesting situation of tight confinement, when the ridge width is of the order of the operating wavelength, the FT distribution with respect to the PBGs is no longer intuitive, and a slight increase of the ridge width can lead also to a counterintuitive increase of the propagation losses. For instance, by increasing the ridge width from $w=1\,\mu\text{m}$ to $w=1.2\,\mu\text{m}$ in case C, the propagation losses increase by almost two orders of magnitude (see Tab.~\ref{tab:cases}).

As a final remark, we stress that, in the above discussion we always considered the situation in which the number $N$ of multilayer periods is sufficiently large so that the confinement is essentially perfect in the whole region defined by the PBG of the corresponding infinite multilayer. For the best structure considered in this Letter, $N=10$ is sufficient to obtain propagation losses that are equivalent to the case of an infinite number of periods. Indeed, in that case, the mode Fourier components are localized well within the PBGs and far from the photonic band edges (see Fig.~\ref{fig:caseC}). However, care should be taken when the Fourier components are nonvanishing nearby the photonic band edge, for a considerably large number of periods could be required to achieve strong attenuation in the multilayer.

In conclusion, we showed that the optimization of the PBGs for both light polarizations is essential to achieve long-range guided BSWs in PhC ridges. The propagation length can be improved by engineering the structure so that the field Fourier components for both polarizations are outside the cladding light cone and inside the corresponding PBG. We illustrated that this can be achieved by a proper design of the multilayer by choosing its period and filling fraction to adjust the PBG position and extension in the Fourier space. Our results clarify the nature of BSW propagation in PhC ridges and extend the possibilities of BSW-based integrated platforms for on-chip light control.

\begin{table}[!b]
\centering
\begin{tabular}{|ccccc|}
\hline
\phantom{case} & $f$ & $w$ ($\mu$m) &\ $n_\text{BSW}$\ \ \ & $\alpha$ (dB/cm) \\ \hline
A & $0.70$ & $1.0$ & $1.444$ & $50.0$\\ \hline
B & $0.45$ & $1.0$ & $1.474$ & $1.03$\\ \hline
C & $0.22$ & $1.0$ & $1.483$ & $5.67\times 10^{-5}$\\
& & $1.2$ & $1.487$ & $5.48\times 10^{-3}$\\ \hline
\end{tabular}
\caption{Results for the cases A, B, and C described in the Letter, with $f$ the filling fraction, $w$ the ridge width, $n_\text{BSW}$ the mode effective index, and $\alpha$ the mode propagation losses.}
\label{tab:cases}
\end{table}

\bibliography{references}

\begin{thebibliography}{17}
\bibitem [{los()}]{loss}%
  \BibitemOpen
  \href@noop {} {}\bibinfo {note} {Here the propagation losses $\alpha$ are
  calculated from the imaginary part of the complex effective index of the mode
  obtained by the FDE solver: $\alpha(\text{dB/m})=\dfrac{4\pi
  n_\text{im}}{\lambda_0}\dfrac{10}{\ln 10}$, where $n_\text{im}$ is the
  imaginary part of the effective index.}\BibitemShut {Stop}%
\end{thebibliography}

\end{document}